\documentclass[conference]{IEEEtran}
\usepackage{blindtext}
\usepackage{amsmath}
\usepackage{amssymb}
\usepackage{bm}
\usepackage{mdwmath}
\usepackage{xfrac}
\usepackage{graphicx}
\usepackage{xcolor}
\usepackage[caption=false,font=footnotesize,subrefformat=parens]{subfig}
\usepackage{makecell}
\usepackage{multirow}
\usepackage{acronym}
\usepackage[noadjust]{cite}
\usepackage{url}
\usepackage{algpseudocode}
\newcounter{MYalgorithmic}
\renewcommand{\theMYalgorithmic}{\arabic{MYalgorithmic}}
\newcommand{\algcaption}[1]{
	\refstepcounter{MYalgorithmic}
	\textbf{Algorithm}~\textbf{\theMYalgorithmic}.~#1}
\newenvironment{MYalgorithmic}[5]
{
	\hrule height 1.2pt
	\vspace{3pt}
	#1{#2}%
	#3{#4}
	\vspace{3pt}
	\hrule height 0.5pt
	\vspace{3pt}
	#5
}
{
	\vspace{3pt}
	\hrule height 0.5pt
}

\newtheorem{definition}{Definition}
\newtheorem{problem}{Problem}
\newtheorem{lemma}{Lemma}

\newcommand{\upperroman}[1]{\uppercase\expandafter{\romannumeral#1}}

\newcommand{\myvec}[1]{\bm{\mathrm{#1}}}
\begin{document}

%\bstctlcite{MyBSTcontrol}
%% *************************************************************************
\title{A Two-Stage Allocation Scheme for Delay-Sensitive Services in Dense Vehicular Networks}

\author{
	\IEEEauthorblockN{Haojun~Yang\IEEEauthorrefmark{1},
		Long~Zhao\IEEEauthorrefmark{1},
		Lei~Lei\IEEEauthorrefmark{2},
		and Kan~Zheng\IEEEauthorrefmark{1}}
	\IEEEauthorblockA{\IEEEauthorrefmark{1}Wireless Signal Processing and Network (WSPN) Lab,\\
		Key Laboratory of Universal Wireless Communication, Ministry of Education,\\
		Beijing University of Posts and Telecommunications (BUPT), Beijing, 100876, China.}
	\IEEEauthorblockA{\IEEEauthorrefmark{2}State Key Lab of Rail Traffic Control and Safety, Beijing Jiaotong University, China. \\
		Email: yanghaojun.yhj@bupt.edu.cn}}

\maketitle

\begin{abstract}
Driven by the rapid development of wireless communication system, more and more vehicular services can be efficiently supported via vehicle-to-everything (V2X) communications. In order to allocate radio resource with the reasonable implementation complexity in dense urban intersection, a two-stage allocation algorithm is proposed in this paper, whose main objective is to minimize delay and ensure reliability. In particular, as for the first stage, the allocation policy is based on traffic density information (TDI), which is different from utilizing channel state information (CSI) and queue state information (QSI) in the second stage. Moreover, in order to reflect the influence of TDI on delay, a macroscopic vehicular mobility model is employed in this paper. Simulation results show that the proposed algorithm can acquire an asymptotically optimal performance with the acceptable complexity.
\end{abstract}

\begin{IEEEkeywords}
Low latency and high reliability, radio resource allocation, dense urban intersection, macroscopic mobility model.
\end{IEEEkeywords}

\IEEEpeerreviewmaketitle

%\renewcommand{\IEEEiedlistdecl}{\IEEEsetlabelwidth{DDDDDDDD}}
%\begin{acronym}
%	\acro{ITS} {Intelligent Transportation System}
%	\acro{LTE} {Long Term Evolution}
%\end{acronym}
%\renewcommand{\IEEEiedlistdecl}{\relax} % remember to reset \IEEEiedlistdecl
%% *************************************************************************

\section{Introduction}
\label{sec:Introduction}

With the rapid development of wireless communication systems, intelligent transportation systems (ITSs) have been widely studied in recent years. More and more vehicular services can be efficiently supported by the evolving wireless networks~\cite{3gpp885,3gpp886}. As a typical dense scenario in vehicular networks, urban intersection is studied in this paper. In order to meet various vehicular requirements, there exist two categories of applications in urban environments, namely non- and delay-sensitive ones~\cite{zhengsurvey2015}. In general, the delay-sensitive services are safety-related, and mainly focus on the performance metrics about low latency and high reliability, such as cooperative driving and road safety, etc. On the other hand, as for non-delay-sensitive services, data rate is a key performance indicator.

%Because of the poor deployment of roadside infrastructures, dedicated short range communication (DSRC) systems are paid less attention in current vehicular networks. Instead, long term evolution (LTE) and its beyond are regarded as the most promising solution to meet various vehicle-to-everything (V2X) communications. Recently, the 3rd generation partnership project (3GPP) declares that LTE-based V2X services adopt PC5, Uu interface and their hybrid to implement information exchange. Utilizing PC5 interface vehicles can directly communicate with other entities via sidelink (SL), which is similar to LTE-based device-to-device (D2D) communications. Obviously, it is more conducive to reduce latency than Uu interface.

Because of the poor deployment of roadside infrastructures, dedicated short range communication (DSRC) systems are paid less attention in current vehicular networks. Instead, long term evolution (LTE) and its beyond are regarded as the most promising solution to meet various vehicle-to-everything (V2X) communications. Recently, the 3rd generation partnership project (3GPP) declares that LTE-based V2X services adopt PC5, Uu interface and their hybrid to implement information exchange.

A theoretical analysis about radio resource management for D2D-based vehicle-to-vehicle (V2V) communication is given in~\cite{Sun2016}, where the scenario of cellular and vehicular users coexistence is studied in detail. However, the characteristics of mobility are not considered in that paper. Although the authors consider traffic model in~\cite{Zheng2016}, their optimization objective is just the delay without paying attention to the reliability. Moreover, their research scenario is focused on the highway. Besides the above work, some other problems about vehicular communications are also studied in~\cite{Zhengtvt2013,Lei2016,Zhang2014,Zhang2015,Zhang2016,Zhengtwc2013,Zhengjsac2013}, such resource allocation and performance analysis, etc.
%~\cite{Ren2015,Lei2016,Huang2016}

Therefore, motivated by the above facts, this paper focuses on the scenario of urban intersection, and aims to investigate radio resource allocation policy to minimize the latency of delay-sensitive services, where the corresponding reliability is considered at the same time. Furthermore, in order to reduce the complexity, a two-stage allocation policy is also proposed, where the allocation based on the traffic density information (TDI) is separately considered. Finally, with the aid of traffic flow theory, we develop a delay utility function adopting macroscopic vehicular mobility model in this paper.

The remainder of this paper is organized as follows. In Section~\ref{sec:System}, the system model and some assumptions are introduced. Section~\ref{sec:Intra} first studies the allocation policy of Stage two based on channel state information (CSI) and queue state information (QSI). Then, Stage one based on TDI, namely inter-subregion resource allocation is discussed in Section~\ref{sec:Inter}. Finally, Section~\ref{sec:Simulation} illustrates the simulation results and conclusions are drawn in Section~\ref{sec:Conclusion}.

%\textit{Notation:} $ \myvec{1}\{ \cdot \} $ denotes the indicator function. $ \lfloor x \rfloor $ represents the largest integer that is smaller than $ x $.

%\textit{Notation:} $ \myvec{1}\{ \cdot \} $ denotes the indicator function. $ \mathbb{E}(\cdot) $ represents mathematical expectation. $ (\cdot)^\text{T} $ denotes the transpose of a matrix or vector. $ \lfloor x \rfloor $ represents the largest integer that is smaller than $ x $. $ \mathbb{CN}(\mu,\sigma^2) $ is the complex Gaussian distribution with mean $ \mu $ and real/imaginary component variance $ \sigma^2/2 $.

\section{System Model}
\label{sec:System}

%In this section, we introduce the system model including scenario, channel model, queue model and three performance metrics for two kinds of services.

\subsection{Scenario Description}
\begin{figure}[!t]
	\centering
	\includegraphics[scale=0.33]{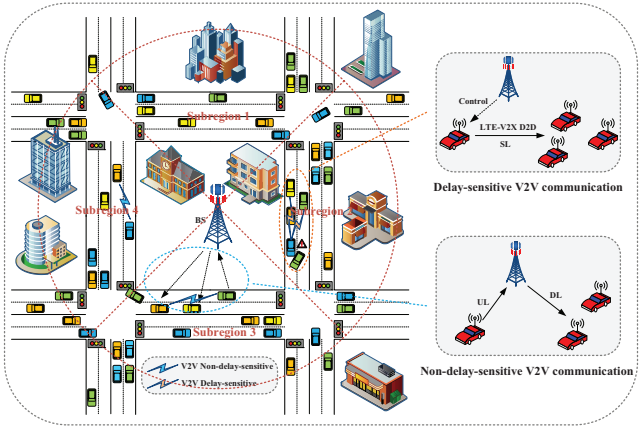}
	\caption{Scenario of urban vehicular networks.}
	\label{fig_1}
\end{figure}

%\begin{figure*}[!t]
%	\centering
%	\subfloat[Urbanuuuuuu6uu scenario.]{\includegraphics[scale=0.4]{Scenario}\label{fig_1a}}
%	\hfil
%	\subfloat[Resource allocation policy.]{\includegraphics[scale=0.42]{Policy}\label{fig_1b}}
%	\caption{System model.}
%	\label{fig_1}
%\end{figure*}

As shown in Fig.~\ref{fig_1}, consider an urban vehicular network with one base station (BS). Assume that each vehicle associating with BS is equipped with one receiving antenna and $ N_\text{T} $ transmitting antennas. There exist two kinds of services in the network, namely non- and delay-sensitive V2V services. As for delay-sensitive V2V services, LTE-based D2D communication is utilized. On the other hand, non-delay-sensitive services can be provided via traditional LTE network. Note that we only pay attention to the uplink (UL) in this paper.

%\begin{figure}[!t]
%	\centering
%	\includegraphics[scale=0.4]{Policy}
%	\caption{Wireless resource allocation policy.}
%	\label{fig_2}
%\end{figure}

In order to efficiently allocate radio resources in dense urban intersection and reduce the complexity, we propose a two-stage allocation policy. As illustrated in Fig.~\ref{fig_1}, the intersection is divided into four subregions. The first stage is to allocate the resources of each subregion based on the corresponding TDI. Here we assume that different subregions use orthogonal resources. The second stage is about the allocation among intra-subregion. In contrast to that of Stage one, Stage two uses reusable resources.

Assume that the number of non- and delay-sensitive vehicles in a subregion are $ N_1 $ and $ N_2 $, respectively. Since the broadcast characteristic of delay-sensitive services, a number of broadcast links are equivalent to one link for simplicity in this paper. Then the total number of links in the subregion is $ N_\text{L}=N_1+N_2 $. Moreover, there are $ N_\text{RB} $ independent resource blocks (RBs) in the subregion. Each link can be allocated at most one RB. Based on the assumption in most existing works~\cite{Lei2014}, the resource allocated to a delay-sensitive link can be reused by at most one non-delay-sensitive link.

\subsection{Channel Model}
The network is assumed to work in slotted time $ t \in \{1,2,\cdots\} $, and we use slot $ t $ to denote the time interval $ [t,t+1) $. Let $ \myvec{H}_{ij}^k (t) = \sqrt{L_{ij}(t)} \myvec{h}_{ij}^k (t) \in \mathbb{C}^{1 \times N_\text{T}} $ denote the CSI matrix from transmitter $ i $ to receiver $ j $ on $ k $-th RB during slot $ t $, where $ L_{ij}(t) $ is the large-scale fading coefficient containing the path loss and shadow, and $ \myvec{h}_{ij}^k (t) $ is the small-scale fading random variable. Assume that the elements of $ \myvec{h}_{ij}^k (t) = [h_{ij1}^k, h_{ij2}^k, \cdots, h_{ijN_\text{T}}^k] $ are independent and identically distributed (i.i.d) complex Gaussian random variables, namely $ h_{ijm}^k \mathop \sim\limits^{\text{i.i.d.}} \mathbb{CN}(0,1) $. Note that $ j=0 $ represents the receiver is BS. At last, let $ \myvec{H}(t)=\{\myvec{H}_{ij}^k (t)\} \in \mathcal{H} $ denote the network CSI at slot $ t $.

\subsection{Queue Model}
Each vehicle maintains one traffic queue with a finite queue length $ N_\text{Q} < \infty $. Let $ Q_i(t) $ denote the QSI (the number of bits) of vehicle $ i $ at the beginning of slot $ t $. Hence, the queue dynamic is given by
\begin{align}
&\mathrel{\phantom{=}}Q_i(t+1) \notag \\
&= \min \left\{ {N_\text{Q}, \max \left\{ {0, Q_i(t)-\mu_i(t)} \right\} +A_i(t)} \right\},
\end{align}
where $ A_i(t) $ denotes the traffic arrival at the end of slot $ t $, and the traffic departure at slot $ t $ is given by $ \mu_i(t) $. We assume that the traffic arrival $ A_i(t) $ is independent w.r.t. $ i $ and i.i.d. over slots obeying a general distribution with mean $ \mathbb{E}[A_i(t)] = \overline{A_i} $. Let $ \myvec{Q}(t)=\{Q_i(t)\} \in \mathcal{Q} $ denotes the network CSI at slot $ t $.

\subsection{Performance Metrics}
Each service has its specific communication requirements in vehicular network. Hence, it is necessary to study the performance metrics of different services. Let $ s_k^{l}(t) $ be the RB allocation at slot $ t $, the value of $ s_k^{l}(t) $ is defined as
\begin{align}
s_k^{l}(t) = \begin{cases}
1, \; \text{$ k $-th RB is allocated to
	link $ l $ at slot $ t $}, \\
0, \; \text{otherwise},
\end{cases}
\end{align}
where $ k \in \{1, 2, \cdots, N_\text{RB}\} $ and $ l \in \{1, 2, \cdots, N_\text{L}\} $.

\subsubsection{Delay-sensitive Service Metric}
As for delay-sensitive services, we first focus on the packet reception ratio (PRR) which is defined in~\cite{3gpp885}. So we have the following definition.
\begin{definition}[Packet Reception Ratio]
Let $ N_i(t) $ denote the number of the neighborhoods of vehicle $ i $ at slot $ t $, then the PRR is defined as the ratio of successful reception among $ N_i(t) $, i.e.,
\begin{align}
p_i(t) &\triangleq \dfrac{1}{N_i(t)}\sum\limits_{j=1}^{N_i(t)} \myvec{1} \left\{ \rho_j^{(i)}(t) \geqslant \rho_{\text{th}} \right\} \notag \\
&= \dfrac{1}{N_i(t)}\sum\limits_{j=1}^{N_i(t)} \myvec{1} \bigg\{ \sum\limits_{k=1}^{N_\text{RB}} \dfrac{s_k^i(t)P_i(t)|\myvec{H}_{ij}^k (t)|^2}{\sigma^2+\sum\limits_{m=1}^{N_1}s_k^m(t)P_m(t)|\myvec{H}_{m0}^k (t)|^2} \notag \\
&\mathrel{\phantom{=}} \geqslant \rho_{\text{th}} \bigg\},
\end{align}
where $ \rho_j^{(i)}(t) $ is the receiving signal-to-interference-plus-noise ratio (SINR) of vehicle $ j $ among $ N_i(t) $, $ P_i(t) $ is the transmit power of vehicle $ i $, and $ \sigma^2 $ is the power of additive white Gaussian noise. Here successful reception is considered as the fact that SINR is greater than or equal to a threshold $ \rho_{\text{th}} $. Specially, the average PRR $ \overline{p_i} $ can be calculated by the following formula, i.e.,
\begin{align}
\label{average}
\overline{p_i} = \limsup_{T \rightarrow \infty} \dfrac{1}{T} \sum_{t=1}^{T} \mathbb{E}^{\Omega} \left[ p_i(t) \right].
\end{align}
\end{definition}

The PRR is a good proxy for reliability. As for delay, we have the following definition.
\begin{definition}[Average Queue Length]
Assume that $ Q(t) $ is a discrete time queue, then the average queue length under a policy $ \Omega $ is given by
\begin{align}
\overline{Q} \triangleq \limsup_{T \rightarrow \infty} \dfrac{1}{T} \sum_{t=1}^{T} \mathbb{E}^{\Omega} \left[ Q(t) \right].
\end{align}
Furthermore, if the average queue length $ \overline{Q}<\infty $, the discrete time queue is strongly stable. A network of queues is stable if all individual queues of the network are stable. Based on the Little's law, we can also calculate the \textit{average delay}.
\end{definition}

\subsubsection{Non-delay-sensitive Service Metric}
With the regard to the non-delay-sensitive services, we mainly focus on the data rate. In order to simplify the communication model, the perfect CSI at the receiver and transmitter are assumed. Therefore, the maximum achievable data rate of vehicle $ i $ at slot $ t $ is given by
\begin{align}
r_i(t) \triangleq B \log \left( 1+\rho_i(t) \right),
\end{align}
where $ B $ denotes the bandwidth of one RB, and $ \rho_i(t) $ can be calculated as
\begin{align}
\rho_i(t) = \sum_{k=1}^{N_\text{RB}}\dfrac{s_k^i(t)P_i(t)|\myvec{H}_{i0}^k (t)|^2}{\sigma^2+\sum\limits_{j=1}^{N_2}s_k^{j}(t)P_j(t) \max\limits_m \left\{ |\myvec{H}_{jm}^k (t)|^2 \right\}},
\end{align}
where $ m \in \{1, 2, \cdots, N_j(t)\} $. Similarly, we can also utilize Equ.~\eqref{average} to calculate the average data rate $ \overline{r_i} $.

\section{Intra-subregion Resource Allocation}
\label{sec:Intra}

%In the following sections, we first study the allocation policy of stage two based on the CSI and QSI, after that stage one is studied. Next, we formulate the problem of intra-subregion resource allocation, and present optimal policy. Furthermore, considering the complexity of implementation, a low complexity algorithm is proposed for stage two.

%In the following sections, we first study the allocation policy of Stage two based on the CSI and QSI, after that Stage one is studied.

\subsection{Resource Allocation Policy}
In general, a resource allocation policy is a mapping function from the system state to the resource allocation actions. A policy is called feasible if the relevant actions satisfy the required constraints. As previously mentioned, our policy of intra-subregion resource allocation satisfies the following constraints, i.e.,
\begin{align}
\sum_{i=1}^{N_1}s_i^k &\leqslant 1,\forall k \in \{1, 2, \cdots, N_\text{RB}\}, \\
\sum_{i=1}^{N_2}s_i^k &\leqslant 1,\forall k \in \{1, 2, \cdots, N_\text{RB}\}, \\
\sum_{k=1}^{N_\text{RB}}s_i^k &\leqslant 1,\forall i \in \{1, 2, \cdots, N_\text{L}\}.
\end{align}

\subsection{Problem Formulation}
In this paper, our main objective is to minimize the latency of delay-sensitive services, while satisfying corresponding reliability requirements and data rate requirements. Thus, we consider the following optimization problem, i.e.,

\begin{problem}[Delay-optimal Policy for Intra-subregion Resource Allocation]
\label{pro1}
Given a set of feasible policies $ \{\Omega\} $, and assuming $ \myvec{r}_\text{th} = [r_{1}^\text{(th)}, r_{2}^\text{(th)}, \cdots, r_{N_1}^\text{(th)}]^\text{T} $ and $ \myvec{p}_\text{th} = [p_{1}^\text{(th)}, p_{2}^\text{(th)}, \cdots, p_{N_2}^\text{(th)}]^\text{T} $ are the minimum data rate of all non-delay-sensitive vehicles and reliability requirements of all delay-sensitive vehicles, the optimization problem is then formulated as
\begin{align}
\label{problem1}
\begin{array}{c l}
\min\limits_\Omega & d_\text{sum}(\Omega) \triangleq \sum\limits_{i=1}^{N_2} \alpha_i d_i(\Omega) \\
\text{s.t.} &
\begin{cases}
\overline{p_i} \geqslant p_{i}^\text{(th)}, \\
\overline{r_j} \geqslant r_{j}^\text{(th)}, \\
\max\limits_i \left\{ \dfrac{\overline{Q_i}}{\overline{A_i}} \right\} \leqslant \min\limits_j \left\{ \dfrac{\overline{Q_j}}{\overline{A_j}} \right\},
\end{cases}
\end{array}
\end{align}
where $ i \in \{ 1, 2, \cdots, N_2 \} $, $ j \in \{ 1, 2, \cdots, N_1 \} $, and $ \alpha_i $ is the positive weighted factor for each delay-sensitive vehicle.
\end{problem}

In general, with the regard to a unichain policy $ \Omega $, the induced Markov chain is ergodic and there is a unique steady state distribution $ \pi(\Omega) $. Hence, we have
\begin{align}
d_i(\Omega) &= \limsup\limits_{T \rightarrow \infty} \dfrac{1}{T} \sum\limits_{t=1}^{T} \mathbb{E}^\Omega \left[ f(Q_i(t)) \right] \notag \\
&= \mathbb{E}^{\pi(\Omega)} \left[ f(\overline{Q_i}) \right],
\end{align}
where $ f(\overline{Q_i}) = \overline{Q_i}/\overline{A_i} $ denotes the average delay.

\subsection{Elements of MDP}
\begin{figure*}[!t]
	\newcounter{mytempeqncnt}
	\setcounter{mytempeqncnt}{\value{equation}}
	\setcounter{equation}{13}
	\normalsize
	\begin{align}
	\label{Lagrangian}
	\mathcal{H}(\myvec{\beta},\myvec{\gamma},\myvec{\eta},\lambda) &= \min\limits_\Omega \mathcal{L}_2(\Omega;\myvec{\beta},\myvec{\gamma},\myvec{\eta},\lambda) \notag \\
	&= \min\limits_\Omega \limsup\limits_{T \rightarrow \infty} \dfrac{1}{T} \sum\limits_{t=1}^{T} \mathbb{E}^\Omega \bigg\{ \sum\limits_{i=1}^{N_2} \left[ \alpha_i f\left( Q_i(t) \right) -\beta_i \left( p_i(t)-p_i^\text{(th)} \right) \right] -\sum\limits_{j=1}^{N_1} \gamma_j \left( r_j(t)-r_j^\text{(th)} \right) \notag \\
	&\mathrel{\phantom{=}} +\lambda \left( \max\limits_i \left\{ \dfrac{\overline{Q_i}}{\overline{A_i}} \right\}-\min\limits_j \left\{ \dfrac{\overline{Q_j}}{\overline{A_j}} \right\} \right) +\sum\limits_{l=1}^{N_\text{L}} \eta_l \myvec{1}\left( Q_l(t)=N_\text{Q} \right) \bigg\} \notag \\
	&= \min\limits_\Omega \limsup\limits_{T \rightarrow \infty} \dfrac{1}{T} \sum\limits_{t=1}^{T} \mathbb{E}^\Omega \left[ g \left( \chi(t), \Omega(\chi(t)),\myvec{\beta},\myvec{\gamma},\myvec{\eta},\lambda \right) \right].
	\end{align}
%	\begin{align}
%	\label{cost}
%	g(\chi(t), \Omega(\chi(t)),\myvec{\beta},\myvec{\gamma},\myvec{\eta},\lambda) &= \sum\limits_{i=1}^{N_2} \left[ \alpha_i f\left( Q_i(t) \right) -\beta_i \left( p_i(t)-p_i^\text{(th)} \right) \right] -\sum\limits_{j=1}^{N_1} \gamma_j \left( r_j(t)-r_j^\text{(th)} \right) \notag \\
%	&\mathrel{\phantom{=}} +\lambda \left( \max\limits_i \left\{ \dfrac{\overline{Q_i}}{\overline{A_i}} \right\}-\min\limits_j \left\{ \dfrac{\overline{Q_j}}{\overline{A_j}} \right\} \right) +\sum\limits_{l=1}^{N_\text{L}} \eta_l \myvec{1}\left( Q_l(t)=N_\text{Q} \right).
%	\end{align}
	\hrulefill
	\vspace*{-10pt}
	\setcounter{equation}{\value{mytempeqncnt}}
\end{figure*}

The optimization problem is formulated as an \textit{infinite horizon average cost constrained} Markov decision process (MDP). In general, MDP is characterized by five elements, i.e., system state space, action space, state transition kernel, average cost function and constraint conditions as follows.
\begin{itemize}
\item System State Space: $ \{ \chi(t) \} = \{ \myvec{H}(t), \myvec{Q}(t) \} \in \mathcal{X}=\mathcal{H} \times \mathcal{Q} $.
\item Action Space: $ \{ \Omega(\chi(t)) \} $, which is a set of unichain
feasible policies under the system state $ \chi(t) $.
\item State Transition Kernel: $ \Pr[\chi(t+1) | \chi(t), \Omega(\chi(t))] $. Since the property of Markov process, we have
\end{itemize}
\begin{align}
&\mathrel{\phantom{=}}\Pr [ \chi(t+1) | \chi(t), \Omega(\chi(t)) ] \notag \\ 
&=  \Pr [ \myvec{H}(t+1) | \chi(t), \Omega(\chi(t)) ] \Pr [ \myvec{Q}(t+1) | \chi(t), \Omega(\chi(t)) ] \notag \\
&= \Pr [ \myvec{H}(t+1) ] \Pr [ \myvec{Q}(t+1) | \chi(t), \Omega(\chi(t)) ].
\end{align}
\begin{itemize}
\item Average Cost Function and Constraint Conditions: They are described in detail at Equ.~\eqref{problem1}.
\end{itemize}

Because of the constraints in Problem~\ref{pro1}, the standard Lagrangian approach is utilized here. Then the constrained MDP can be transformed to the \textit{unconstrained} MDP, and the Lagrange dual function is also defined as Equ.~\eqref{Lagrangian} listed at the top of this page, where $ \myvec{\beta}=\{\beta_i \geqslant 0\} $, $ \myvec{\gamma}=\{\gamma_j \geqslant 0\} $, $ \myvec{\eta}=\{\eta_l \geqslant 0\} $  and $ \lambda \geqslant 0 $ are the Lagrange multipliers. Therefore, the average cost function of the corresponding unconstrained MDP can be obtained from Equ.~\eqref{Lagrangian}. As a rule, the delay-optimal policy can be obtained by solving the Bellman equation~\cite{Dimitribook}, we discuss it in the next subsection.

\subsection{Optimal Solution of MDP}
As previously mentioned, we have converted Problem~\ref{pro1} into the unconstrained MDP, thus it can be solved by Bellman equation expressed as follows.
\begin{lemma}[Bellman Equation]
For any given $ \myvec{\beta} $, $ \myvec{\gamma} $, $ \myvec{\eta} $ and $ \lambda $, if there exist a scalar $ \theta $ and a vector $ \myvec{V}=\left[ V(\chi^1), V(\chi^2), \cdots \right]  $ satisfy the Bellman equation for the delay-optimal unconstrained MDP in Equ.~\eqref{Lagrangian}, namely
\setcounter{equation}{14}
\begin{align}
\label{Bellman}
&\mathrel{\phantom{=}}\theta+V(\chi^i) \notag \\
&=\min\limits_{\Omega(\chi^i)} \Big\{ g(\chi^i, \Omega(\chi^i),\myvec{\beta},\myvec{\gamma},\myvec{\eta},\lambda) \notag \\
&\mathrel{\phantom{=}}+\sum_{\chi^j}\Pr \left[ \chi^j|\chi^i,\Omega(\chi^i) \right]V(\chi^j) \Big\}, \forall \chi^i \in \mathcal{X},
\end{align}
then $ \theta=\min_\Omega \mathcal{L}(\Omega,\myvec{\beta},\myvec{\gamma},\myvec{\eta},\lambda) $ is the optimal average cost per-stage, and the optimal policy for Problem~\ref{pro1} is $ \Omega^* $, which minimizes the R.H.S. of Equ.~\eqref{Bellman} for any state $ \chi^i \in \mathcal{X} $. Similarly, as for a unichain policy, there is a unique solution to Equ.~\eqref{Bellman}. Therefore, we only consider the unichain feasible policy in this paper.
\end{lemma}

It is well known that the system state space gradually becomes huge with the increasing number of vehicles. Therefore, in order to reduce the complexity, the reduced-state Bellman equation can be adopted to solve Problem~\ref{pro1}~\cite{Lau2010}, which only takes advantage of the QSI. Then we have the following lemma, i.e.,
\begin{lemma}[Reduced-State Bellman Equation]
%In general, solving the original and reduced-state Bellman equation own the equivalent results.
In general, the equation can be given by
\begin{align}
\label{Reduced_Bellman}
&\mathrel{\phantom{=}}\theta+\tilde V(\myvec{Q}^i) \notag \\
&=\min\limits_{\Omega(\myvec{Q}^i)} \Big\{ \tilde g(\myvec{Q}^i, \Omega(\myvec{Q}^i),\myvec{\beta},\myvec{\gamma},\myvec{\eta},\lambda) \notag \\
&\mathrel{\phantom{=}}+\sum_{\myvec{Q}^j} \tilde f \left( \myvec{Q}^j|\myvec{Q}^i,\Omega(\myvec{Q}^i) \right) \tilde V(\myvec{Q}^j) \Big\}, \forall \myvec{Q}^i \in \mathcal{Q},
\end{align}
where $ \tilde V(\myvec{Q})=\mathbb{E}[V(\chi)|\myvec{Q}] $, $ \tilde g(\myvec{Q}, \Omega(\myvec{Q}),\myvec{\beta},\myvec{\gamma},\myvec{\eta},\lambda)=\mathbb{E}[g(\chi, \Omega(\chi),\myvec{\beta},\myvec{\gamma},\myvec{\eta},\lambda)|\myvec{Q}] $ and $ \tilde f (\myvec{Q}^j|\myvec{Q}^i,\Omega(\myvec{Q}^i))=\mathbb{E}[\Pr (\myvec{Q}^j|\chi^i,\Omega(\chi^i)) | \myvec{Q}^i] $ are conditional potential function, average cost per-stage and average transition
kernel, respectively.
\end{lemma}

\section{Inter-subregion Resource Allocation}
\label{sec:Inter}

\subsection{Fundamentals of Traffic Flow Theory}
According to the different traffic characteristics, vehicular mobility models are usually classified into two categories, namely macroscopic and microscopic models. Each category of model focuses on different performance indicators. The macroscopic models generally describe the average behavior of many vehicles at specific location and time, treating traffic flow as fluid dynamics. Therefore, vehicular density and mean velocity are considered in the macroscopic models, which raises the traffic flow theory. However, the microscopic models describe the precise behavior of each system entity (i.e., vehicle or driver), hence they are more complicated than the macroscopic models.

In order to allocate wireless resources efficiently among inter-subregion, we model the TDI adopting the traffic flow theory. It is well known that there are many macroscopic models, such as Greenshield's model, Greenberg's model, Underwood's model, etc. For the sake of simplicity, we utilize linear Greenshield's model in this paper. Here we give a brief introduction about Greenshield's model. In general, there exist two parameters in the Greenshield's model, namely free flow speed $ v_\text{free} $ and jam density $ \kappa_\text{jam} $~\cite{trafficflow}. The relationship between flow $ f $ and density $ \kappa $ is given by
\begin{align}
\label{density}
f = \kappa v_\text{free} - \dfrac{\kappa^2}{\kappa_\text{jam}}v_\text{free}.
\end{align}
%\begin{align}
%\label{speed}
%v = v_\text{free} \left( 1-\dfrac{\kappa}{\kappa_\text{jam}} \right).
%\end{align}
%Furthermore, we have flow $ f = \kappa v $. Substituting Equ.~\eqref{speed} into this, then we get the relationship between flow and density, i.e.,
%%\begin{align}
%%\label{flow}
%%f = \kappa v,
%%\end{align}
%\begin{align}
%\label{density}
%f = \kappa v_\text{free} - \dfrac{\kappa^2}{\kappa_\text{jam}}v_\text{free}.
%\end{align}

\begin{figure}[!t]
	\centering
	\includegraphics[scale=0.24]{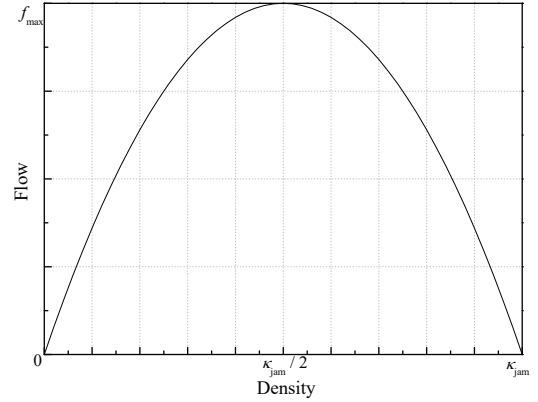}
	\caption{Illustration of density and flow in the Greenshield's model.}
	\label{fig_3}
\end{figure}

We plot Equ.~\eqref{density} in Fig.~\ref{fig_3} to illustrate this relationship. As we see, Fig.~\ref{fig_3} clearly illustrates that the flow increases with the increasing density when $ \kappa \leqslant \kappa_\text{jam}/2 $. It just is a simple parabola.

\subsection{Delay Utility Function}
%As previously mentioned, the allocation policy in the first stage is based on the TDI of corresponding subregion. Therefore, in order to reflect the influence of TDI on delay-sensitive services, we construct a delay utility function with the help of the Greenshield's model. As illustrated in Fig.~\ref{fig_3}, the utility function should satisfy the following properties, i.e.,
In order to reflect the influence of TDI on delay-sensitive services, we construct a delay utility function with the help of the Greenshield's model. As illustrated in Fig.~\ref{fig_3}, the utility function should satisfy the following properties, i.e.,
\begin{itemize}
\item When $ \kappa_i \leqslant \kappa_\text{jam}/2 $, the flow increases with the increase of density, hence the number of delay-sensitive services increases, and the delay requirement gradually increases; and
\item When $ \kappa_i > \kappa_\text{jam}/2 $, the flow decreases with the increase of density, hence the delay requirement gradually decreases for the same reason; and
\item Furthermore, no matter how much the value of $ \kappa_i $, the delay requirement is not equal to zero. Meanwhile, the requirement is normalized for the sake of simplicity; and
\item Let $ \varepsilon_i $ denote the ratio of allocation for subregion $ i $. For any $ \varepsilon_i \in [0,1] $, the allocation efficiency increases with the increase of $ \varepsilon_i $.
\end{itemize}

In conclusion, the utility function is given by
\begin{align}
\label{utility}
U_i(\kappa_i,\varepsilon_i)=\exp\left( -\dfrac{(\kappa_i-\frac{\kappa_\text{jam}}{2})^2}{c_1} \right) \log \left( 1+c_2\varepsilon_i \right),
\end{align}
%\begin{align}
%\label{utility}
%U_i(\kappa_i,\varepsilon_i)=\log \left( 1+\dfrac{e^{-\frac{(\kappa_i-\frac{\kappa_\text{jam}}{2})^2}{c_1}}}{e^{-c_2\varepsilon_i}} \right),
%\end{align}
where $ c_1,c_2>0 $ are constants, which is related to the practical traffic condition. The logarithmic utility function can ensure the fairness, and thus is employed in this paper. In Equ.~\eqref{utility}, the first and second terms represent the normalized delay requirement and the allocation efficiency, respectively. Note that the utility function is just a proxy for delay, not the true value.

\subsection{Problem Formulation}
Comparing to the CSI and QSI in Stage two, the TDI in Stage one changes at a longer time-scale. Therefore, we can formulate a new problem independent of Problem~\ref{pro1}. The main objective of Stage one is to maximize the sum of delay utility defined in Equ.~\eqref{utility} based on the corresponding TDI. The following optimization problem is considered, i.e.,
\begin{problem}[Delay-optimal Policy for Inter-subregion Resource Allocation]
\label{pro2}
Given the TDI of four subregions $ \myvec{\kappa}=[\kappa_1,\kappa_2,\kappa_3,\kappa_4]^\text{T} $, the utility maximization problem of Stage one is then formulated as
\begin{align}
\label{problem2}
\begin{array}{c l}
\max\limits_{\myvec{\varepsilon}} & U_\text{sum}(\myvec{\varepsilon}) \triangleq \sum\limits_{i=1}^{4} U_i(\kappa_i,\varepsilon_i) \\
\text{s.t.} &
\begin{cases}
\varepsilon_i \geqslant 0, \\
\sum\limits_{i=1}^{4} \varepsilon_i = 1.
\end{cases}
\end{array}
\end{align}
%This utility maximization problem is solved in the next subsection.
\end{problem}

\subsection{Resource Allocation for Stage One}
Since the logarithmic function is convex, the compound utility function is convex. We can solve Problem~\ref{pro2} utilizing convex optimization theory~\cite{Boydbook}. First of all, we write the Lagrange function of Problem~\ref{pro2} as follows.
\begin{align}
\mathcal{L}_1(\myvec{\varepsilon};\myvec{\delta},\omega) = -\sum_{i=1}^{4} U_i(\kappa_i,\varepsilon_i) - \sum_{i=1}^{4} \delta_i\varepsilon_i + \omega\left( \sum_{i=1}^{4} \varepsilon_i-1 \right),
\end{align}
where $ \myvec{\delta}=\{\delta_i \geqslant 0\} $ and $ \omega $ are the Lagrange multipliers. Therefore, based on the Karush-Kuhn-Tucker (KKT) condition, we get
\begin{align}
\begin{aligned}
\varepsilon_i \geqslant 0, \quad \sum\limits_{i=1}^{4} \varepsilon_i = 1, \quad \delta_i \geqslant 0, \quad \delta_i\varepsilon_i &=0, \\
\dfrac{\partial \mathcal{L}_1(\myvec{\varepsilon};\myvec{\delta},\omega)}{\partial \varepsilon_i} =-\dfrac{\exp(-\frac{(\kappa_i-\frac{\kappa_\text{jam}}{2})^2}{c_1}) c_2}{1+c_2\varepsilon_i} -\delta_i+\omega &=0,
\end{aligned}
\end{align}
where $ i \in \{ 1,2,3,4 \} $. Then, the utility maximization resource allocation can be given by
%\begin{align}
%\begin{aligned}
%\varepsilon_i & \geqslant 0, \\
%\sum\limits_{i=1}^{4} \varepsilon_i &=1, \\
%\omega -\dfrac{\exp(-\frac{(\kappa_i-\frac{\kappa_\text{jam}}{2})^2}{c_1}) c_2}{1+c_2\varepsilon_i} & \geqslant 0, \\
%\left( \omega -\dfrac{\exp(-\frac{(\kappa_i-\frac{\kappa_\text{jam}}{2})^2}{c_1}) c_2}{1+c_2\varepsilon_i} \right) \varepsilon_i &=0.
%\end{aligned}
%\end{align}
%\begin{align}
%\varepsilon_i =
%\begin{cases}
%\dfrac{1}{c_2} \left( \dfrac{T_ic_2}{\omega}-1 \right), \quad &\omega < T_ic_2, \\
%0, \quad &\omega \geqslant T_ic_2.
%\end{cases}
%\end{align}
\begin{align}
\label{solution}
\varepsilon_i = \max \left\{ 0,\dfrac{1}{c_2} \left( \dfrac{c_2}{\omega} \exp\left( -\frac{(\kappa_i-\frac{\kappa_\text{jam}}{2})^2}{c_1} \right) -1 \right) \right\},
\end{align}
where the Lagrange multiplier $ \omega $ is determined by equation $ \sum_{i=1}^{4} \varepsilon_i = 1 $.

\subsection{Resource Allocation Algorithm for Urban Vehicular Network}
According to the above discussions, we summarize each step of resource allocation for urban vehicular network in detail at Alg.~\ref{alg1}. In Alg.~\ref{alg1}, the Lagrange thresholds $ \overline{\omega_i} $ can be calculated by
\begin{align}
\label{thresholds1}
\overline{\omega_i} = c_2\exp\left( -\frac{(\kappa_i-\frac{\kappa_\text{jam}}{2})^2}{c_1} \right),
\end{align}
and $ \omega_m $ can be calculated by
\begin{align}
\label{thresholds2}
\omega_m = \dfrac{\sum\limits_{i=1}^{m}\exp( -\frac{(\kappa_i-\frac{\kappa_\text{jam}}{2})^2}{c_1} )}{1+\frac{m}{c_2}}.
\end{align}

\begin{figure}[!t]
	\begin{MYalgorithmic}
		\algcaption{Resource allocation algorithm for urban vehicular network}
		\label{alg1}
		\begin{algorithmic}
			\renewcommand{\algorithmicrequire}{\textbf{Initialization:}}
			\Require
			\State \labelitemi~There are a total of $ N_\text{RB}^\text{total} $ RBs at the BS.
			\State \labelitemi~BS gathers the periodic TDI $ \myvec{\kappa}=[\kappa_1,\kappa_2,\kappa_3,\kappa_4]^\text{T} $ from the traffic monitor nodes.
			\State \labelitemi~All vehicles send the QSI to BS.
			\State \labelitemi~BS sets the initial number of subregions $ M=0 $, and the Lagrange multiplier $ \omega=\infty $.
			\renewcommand{\algorithmicrequire}{\textbf{Step 1: Resource allocation for Stage one}}
			\Require
			\State \labelitemi~Calculate the Lagrange thresholds $ \overline{\omega_i} \; (i=1,2,3,4) $ based on Equ.~\eqref{thresholds1}.
			\State \labelitemi~Sort the thresholds with descending order, and obtain the vector $ [ \overline{\omega_{(1)}},\overline{\omega_{(2)}},\overline{\omega_{(3)}},\overline{\omega_{(4)}} ]^\text{T} $.
			\Loop \quad $ m = 1 \rightarrow 4 $
			\State \textbf{1)} Calculate $ \omega_m $ according to Equ.~\eqref{thresholds2}.
			\State \textbf{2)} \textbf{If} $ \omega_m < \overline{\omega_{(m)}} $, \textit{\textbf{let}} $ M=m $, $ \omega=\omega_m $ and \textit{\textbf{continue}}; \textbf{Else} \textit{\textbf{break}}.
			\EndLoop
			\State \labelitemi~Calculate the ratio of allocation $ \varepsilon_i \; (i=1,2,3,4) $ for each subregion based on Equ.~\eqref{solution} with the obtained $ \omega $.
			\renewcommand{\algorithmicrequire}{\textbf{Step 2: Resource allocation for Stage two}}
			\Require
			\State \labelitemi~Calculate the number of RBs for each subregion $ N_\text{RB}^{(i)}= \lfloor \varepsilon_iN_\text{RB}^\text{total} \rfloor $.
			\State \labelitemi~At the beginning of each scheduling slot, execute the algorithm described in Equ.~\eqref{Reduced_Bellman} for each subregion according to the corresponding QSI.
			\State \labelitemi~\textbf{If} the current slot is the moment of periodic TDI report, \textit{\textbf{go to Step 1}} and \textit{\textbf{continue}}; \textbf{Else} \textit{\textbf{loop Step 2}}.
		\end{algorithmic}
	\end{MYalgorithmic}
\end{figure}

\section{Simulation Results}
\label{sec:Simulation}

In order to evaluate the performance of the proposed allocation algorithm, part of the simulation results are shown in this section. For the purpose of better illustration, some simulation assumptions are summarized in Table~as follows.

\begin{table}[!t]
	\renewcommand{\arraystretch}{1.2}
	\setlength{\extrarowheight}{1pt}
	\centering
	\caption{Simulation parameters.}
	\begin{tabular}{ l | l }
		\hline
		\textbf{Parameter} & \textbf{Assumption} \\
		\hline
		\hline
		Bandwidth & 5 MHz \\
		\hline
		$ N_\text{T} $ & 2 transmitting antennas \\
		\hline
		Average packet size & \makecell*[l]{20 bytes for delay-sensitive services \\ 300 bytes for non-delay-sensitive ones} \\
		\hline
		Average arrival rate & 5:5:30 packets/s \\
		\hline
		Queue size & 10 packets \\ 
		\hline
		Scheduling slot & 1 ms (one slot in LTE) \\
		\hline
		TDI update interval & 500 ms \\
		\hline
		$ \kappa_\text{jam} $ & 2 \\
		\hline
	\end{tabular}
	\label{table_1}
\end{table}

\begin{figure}[!t]
	\centering
	\includegraphics[scale=0.29]{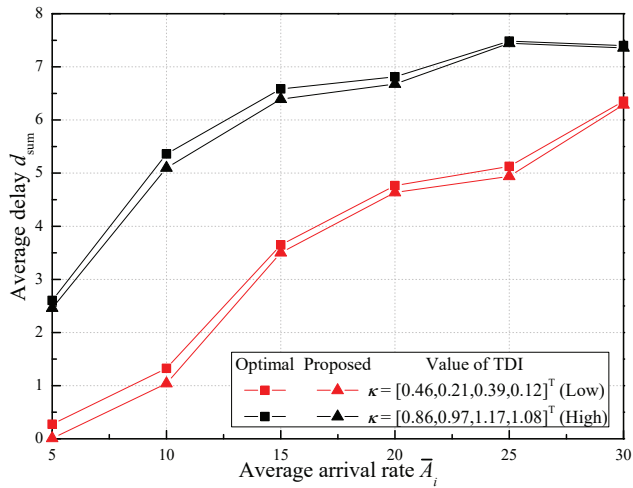}
	\caption{Delay performance versus various arrival rates.}
	\label{fig_sim}
\end{figure}

Fig.~\ref{fig_sim} illustrates that the performances of average delay versus average arrival rate. As can be seen from Fig.~\ref{fig_sim}, the average delay increases with the increasing TDI $ \myvec{\kappa} $, where the high and low TDI are generated by Uniform(0,0.5) and Uniform (0.8,1.2), respectively. Moreover, we also find that the optimal policy solved by the original Bellman equation has the best delay performance at the expense of high implementation complexity. As for the proposed algorithm, it acquires an asymptotically optimal performance, but its complexity has a significant decrease, which is very satisfactory. In particular, when $ \overline{A_i}=25 $ packets/s, the proposed algorithm has the approximately equal performance with the optimal one in the case of high TDI.

\section{Conclusion}
\label{sec:Conclusion}

%In order to reduce the allocation complexity in dense urban intersection, this paper proposed a two-stage allocation algorithm, where stage one utilized the TDI of corresponding subregion to maximize the delay utility. However, as for stage two, the main optimization objective was to minimize the latency of delay-sensitive services, while satisfying the corresponding reliability requirements and data rate requirements. Taking advantage of the Greenshield's model, we also developed a delay utility function to accurately reflect the influence of traffic on latency. Finally, comparing to the optimal solution of MDP, we illustrated that the proposed scheme can reduce the complexity and ensure the performance.

In order to reduce the allocation complexity in dense urban intersection, this paper proposed a two-stage allocation algorithm, where Stage one utilized the TDI of corresponding subregion to maximize the delay utility. While for Stage two, its main optimization objective was to minimize the latency of delay-sensitive services, meanwhile satisfying the corresponding reliability requirements and data rate requirements. Finally, comparing to the optimal solution of MDP, simulation results illustrated that the proposed scheme can acquire an asymptotically optimal performance with the reasonable complexity comparing to the optimal one.

%% *************************************************************************
\section*{Acknowledgment}
\addcontentsline{toc}{section}{Acknowledgment}

This work is supported by the National High Technology Research and Development Program of China under Grant 2014AA01A705, the ``Research and evaluation on key technologies of 5G mobile wireless transmission'', Ministry of Education--China Mobile Research Foundation under Grant MCM20150101, the National Key Scientific Instrument and Equipment Development Project under under Grant 2013YQ20060706, the National Natural Science Foundation of China under Grant 61331009.
%% *************************************************************************
%% References section
%%
%% Can use a bibliography generated by BibTeX as a .bbl file.
%%
%\bibliographystyle{IEEEtran}
%% Argument is your BibTeX string definitions and bibliography database(s).
%\bibliography{IEEEabrv,../Bib/Ref}
%%
%% <OR> Manually copy in the resultant .bbl file.
%% Set second argument of \begin to the number of references.
%% (used to reserve space for the reference number labels box)
%%

%%
%% *************************************************************************
%% End All
\end{document}